\documentclass[preprint,12pt]{elsarticle}

 \usepackage{graphicx}
\usepackage{amssymb}
\usepackage{lineno}

\biboptions{sort&compress}

\usepackage{booktabs}

\newcommand{\pt}{\ensuremath{p_{\mathrm{T}}}}
\newcommand{\ptmax}{\ensuremath{p_{\mathrm{T}}^{\mathrm{max}}}}
\newcommand{\mtop}{\ensuremath{m_{\mathrm{top}}}}
\newcommand{\mW}{\ensuremath{m_{\mathrm{W}}}}
\newcommand{\gtop}{\ensuremath{\Gamma_{\mathrm{top}}}}
\newcommand{\gW}{\ensuremath{\Gamma_{\mathrm{W}}}}
\newcommand{\met}{\ensuremath{E_{T}^{\mathrm{miss}}}}
\newcommand{\mex}{\ensuremath{E_{x}^{\mathrm{miss}}}}
\newcommand{\mey}{\ensuremath{E_{y}^{\mathrm{miss}}}}
\newcommand{\mexy}{\ensuremath{E_{x/y}^{\mathrm{miss}}}}
\newcommand{\ttbar}{\ensuremath{t\bar{t}}}

\journal{Nuclear Instruments and Methods A}

\begin{document}

\begin{frontmatter}



\title{A likelihood-based reconstruction algorithm for top-quark pairs and the KLFitter framework}


\author[Goe,Yale]{Johannes Erdmann}
\author[Goe,Albany]{Stefan Guindon}
\author[Goe]{Kevin Kr{\"o}ninger}
\author[Goe]{Boris Lemmer}
\author[Goe]{Olaf Nackenhorst}
\author[Goe]{Arnulf Quadt}
\author[Goe]{Philipp Stolte}
\address[Goe]{II. Physikalisches Institut, Georg-August-Universit{\"a}t, G{\"o}ttingen, Germany}
\address[Yale]{Now at: Department of Physics, Yale University, New Haven, CT, United States of America}
\address[Albany]{Now at: Department of Physics, SUNY Albany, Albany, NY, United States of America}

\begin{abstract}
A likelihood-based reconstruction algorithm for arbitrary event
topologies is introduced and, as an example, applied to the
single-lepton decay mode of top-quark pair production. The algorithm
comes with several options which further improve its performance, in
particular the reconstruction efficiency, i.e., the fraction of events
for which the observed jets and leptons can be correctly associated
with the final-state particles of the corresponding event
topology. The performance is compared to that of well-established
reconstruction algorithms using a common framework for kinematic
fitting. This framework has a modular structure which describes the
physics processes and detector models independently. The implemented
algorithms are generic and can easily be ported from one experiment to
another.
\end{abstract}


\end{frontmatter}

%\linenumbers

\section{Introduction}
\label{sec:introduction}

Top quarks are produced abundantly at the LHC. Their production
mechanisms as well as their properties are the focus of intensive
studies, exploiting the excellent performance of the ATLAS and CMS
detectors.

Due to the short lifetime of top quarks, their properties can only be
studied indirectly based on their decay products and their
corresponding signatures in the detector, i.e. jets, charged leptons
and missing transverse momentum. The full reconstruction of the
top-quark four-momenta is necessary for a number of precision
measurements, for example measurements of the top-quark mass~\cite{ATLAS:2012aj,CMS:2012tm} and of
differential distributions~\cite{D0:2011wh,CDF:2013tas}. 
It can also be beneficial for searches for
rare processes involving top quarks, for example $\ttbar
H$-production. However, jets cannot be associated
uniquely to the partons of the hard-scattering process and hence
reconstruction algorithms are used to find the best corresponding
match between them. Inefficiencies in these reconstruction algorithms
result in \emph{combinatorial background} which may decrease the
precision of a measurement. %(H\rightarrow \bbbar$)

The aim of this publication is to introduce a likelihood-based method
for kinematic fitting and to compare its performance to that of
alternative reconstruction algorithms. Reconstruction efficiencies and
properties of the reconstructed objects are studied based on a sample
of simulated top-quark pairs produced at a proton-proton collider at a
center-of-mass energy of $\sqrt{s}=7$\,TeV and decaying to a final
state containing exactly one electron. Muons are not considered for reasons
of simplicity. The detector response is simulated by smearing the particle 
energies with assumed resolution functions. All reconstruction algorithms
are implemented in a common framework for kinematic fitting, the
\emph{Kinematic Likelihood Fitter} (\emph{KLFitter}).

Section~\ref{sec:algorithms} describes the event signature of the
studied process and introduces three reconstruction
algorithms. The KLFitter framework is briefly described in
Section~\ref{sec:klfitter}. The Monte Carlo sample used in this study
and the detector modeling applied are presented in
Section~\ref{sec:mc}, followed by the event selection in
Section~\ref{sec:selection}. The performances of the different
reconstruction algorithms are compared in
Section~\ref{sec:performance}. The last section concludes the paper
and gives an outlook on the performance of kinematic fitting of
top-quark pairs produced at a center-of-mass energy of
$\sqrt{s}=14$\,TeV.

\section{Event signatures and reconstruction algorithms}
\label{sec:algorithms}

\begin{figure}[ht!]
\begin{center}
\includegraphics[width=0.50\textwidth]{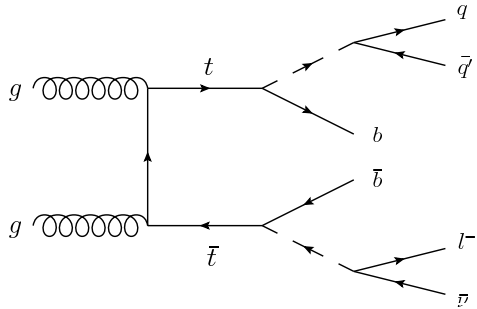}
\caption{Example of a leading-order Feynman diagram of top-quark pair
production and the subsequent decay in the single-lepton decay
mode~\cite{Binosi:2008ig}.
\label{fig:feynman}}
\end{center}
\end{figure}

Top quarks decay to a $W$ boson and a bottom quark in nearly 100\% of
all cases. Consequently, the final state of a top-quark pair is
characterized by the decay products of the two $W$ bosons. If one of
the $W$ bosons decays into a charged lepton and a neutrino while the
other one decays into a pair of quarks, the decay mode is referred to
as the \emph{single-lepton decay mode}. The fraction of top-quark
pairs decaying either in the single-electron or single-muon decay mode
is about 30\%. The corresponding event signature is defined by exactly
one electron or muon, four jets out of which two contain a $B$~hadron,
and a large amount of missing transverse momentum due to the
undetected neutrino. Figure~\ref{fig:feynman} shows an example of a
leading-order Feynman diagram of top-quark pair production and the
subsequent decay in the single-lepton decay mode. 

In the following, we will only refer to quarks in the context of the
hard-scattering process, or on the \emph{parton level}. These quarks will
hadronize and build showers of stable particles, which we will refer
to as the \emph{particle jets} or the \emph{particle level}. It is
naively expected that the four quarks on the parton level will result in
four particle-level jets. However, additional radiation and jet
merging can lead to more or less jets, respectively. Detector effects
can further reduce the number of jets observable in a detector, or on the
\emph{reconstruction level}. In hadron colliders, additional interactions 
(pile-up) can increase the number of reconstructed jets.
The aim of the reconstruction algorithms is to find the correct association 
between the reconstructed jets and the particle-level jets.

In the current example, only four jets are used for the reconstruction
of the top-quark pair. Other jets, e.g., from initial- and final-state
radiation, are ignored (see Section~\ref{sec:selection} for further
details of the event selection). This results in $4!=24$ possible ways
to associate a reconstructed jet with a particle-level jet, referred
to as \emph{permutations} in the following. None of the three
algorithms described in this paper is sensitive to a commutation 
of the two jets from the hadronically decaying $W$ boson, so the number of
distinguishable permutations reduces to 12.

The likelihood-based reconstruction method is described in the
following as well as the two commonly used algorithms, the
\ptmax-method~\cite{Aad:2009wy,thesis_weigell} and the
$\chi^{2}$-method~\cite{Snyder:1995hg}. While the former
algorithm was used to estimate the sensitivity of several ATLAS
measurements, variants of the latter are currently in use at Tevatron
and LHC experiments. 
All three algorithms are based on a simplified leading-order
picture of the physics processes in which no distinction is made
between the particle-level jets and the parton-level quarks. In
addition, the $\chi^{2}$- and likelihood methods include terms
which describe the mapping between particle-level and
reconstructed jets, i.e. taking into account detector resolution
effects. Details on the detector model are given in Section~\ref{sec:mc}.

\subsection{The \ptmax-method}

For a given set of three (out of the four) jets, the transverse
momentum of the three-jet system is calculated. The set with the
largest transverse momentum, $\pt^{\mathrm{max}}$, is associated with
the decay products of the hadronically decaying top quark while the
remaining jet is identified with the $b$-jet from the leptonically
decaying top quark.

\subsection{The $\chi^{2}$-method}

A $\chi^{2}$ variable\footnote{Several definitions for such a
  $\chi^{2}$ variable can be found in the literature, some of which
  are solved iteratively using Lagrangian
  multipliers~\cite{Lyons}. The version chosen in this paper is
  closely related to the likelihood defined in
  Section~\ref{sec:lhmethod}.} is minimized for each permutation of
  jets, defined as
\begin{linenomath}
\begin{eqnarray}
\chi^{2} & = & 8 \cdot \ln 2 \cdot \frac{(m_{q_{1}q_{2}q_{3}} - \mtop)^{2}}{\gtop^{2}}
         + 8 \cdot \ln 2 \cdot \frac{(m_{q_{1}q_{2}} - \mW)^{2}}{\gW^{2}} \nonumber \\
& + &      8 \cdot \ln 2 \cdot \frac{(m_{q_{4}\ell \nu}   - \mtop)^{2}}{\gtop^{2}}
         + 8 \cdot \ln 2 \cdot \frac{(m_{\ell \nu}   - \mW)^{2}}{\gW^{2}} \nonumber \\
& - &  2 \cdot      \sum_{i=1}^{4} \ln W_{\mathrm{jet}}(E^{\mathrm{meas}}_{\mathrm{jet},i} | E_{\mathrm{jet},i}) - 2 \cdot\ln W_{\mathrm{\ell}}(E^{\mathrm{meas}}_{\ell} | E_{\ell}) \nonumber \\
 & - & 2 \cdot\ln W_{\mathrm{miss}}(\mex | p_{x}^{\nu}) - 2 \cdot\ln W_{\mathrm{miss}}(\mey | p_{y}^{\nu}) \, ,
\end{eqnarray}
\end{linenomath}
and the permutation with the smallest $\chi^{2}$ is chosen as an
estimate of the correct association of the jets to the final-state
particles. The free parameters of the $\chi^{2}$ are the mass of the
top quark, \mtop, the particle-jet energies, $E_{\mathrm{jet},i}~(i=1,\dots,4)$,
and the energy of the charged lepton, $E_{\ell}$. 
The first (second) two terms in the first (second) row represent 
the Gaussian constraints on the hadronic
(leptonic) decay branch of the event, where $m_{\mathrm{W}}=80.4\,\mathrm{GeV}$
is the mass of the $W$ boson, and $\Gamma_{\mathrm{top}}=1.5\,\mathrm{GeV}$ and
$\Gamma_{\mathrm{W}}=2.1\,\mathrm{GeV}$ are the decay widths of the top quark
and the $W$ boson, respectively.\footnote{The prefactors of the first
four terms are due to the conversion of the full width at half maximum
into standard deviations: $\Gamma = 2\sqrt{2 \ln 2} \cdot \sigma$.} %The decay width of the top quark is calculated as a function of the top-quark mass according to Ref.~\cite{Oliveira:2001vw}. 
The expressions $m_{q_{1}q_{2}q_{3}}$,
$m_{q_{1}q_{2}}$, $m_{q_{4}\ell \nu}$ and $m_{\ell \nu}$ are invariant
masses calculated from the reconstructed particles' four-momenta. The
last terms in the $\chi^{2}$ constrain the particle energies based on
the measured energies, $E^{\mathrm{meas}}_{\mathrm{jet},i}~(i=1,\dots,4)$ and
$E^{\mathrm{meas}}_{\ell}$. These terms are referred to as
\emph{transfer functions} which in this case are assumed to have a
Gaussian shape (see Section~\ref{sec:mc} for a detailed
discussion). Additionally, $W_{\mathrm{miss}}(\mexy | p_{x/y}^{\nu})$ is the 
transfer function for the $x$- and $y$-components of the 
missing transverse momentum, \met, and
the components of the neutrino momentum. The angles of the particle
jets and the charged lepton are assumed to be measured with negligible
uncertainty.  

Two different options of the $\chi^{2}$-method concerning the treatment of 
the components of the neutrino momentum are tested. 
For the first option, $p_{x}^{\nu}$ and $p_{y}^{\nu}$ are determined from the
transverse momentum balance of all considered particles in a leading-order approach, 
thus constraining the transverse momentum of the top-quark pair to zero.
The $z$-direction of the neutrino momentum, $p_{z}^{\nu}$, is calculated 
using a constraint on the $W$-boson mass, i.e., solving the equation
\begin{linenomath}
\begin{equation}
\label{eqn:pznu}
m_{W}^{2} = \left( p_{\nu} + p_{\ell} \right)^{2}
\end{equation}
\end{linenomath}
for $p_{\nu}^{z}$. Here, $p_{\nu}$ and $p_{\ell}$ are the four-momenta
of the neutrino and the charged lepton, respectively. In case two
solutions exist, the one with the smaller $\chi^{2}$ is chosen. If no
solution exists, the longitudinal momentum is set to $0$\,GeV. 

The second option treats the momentum components of the neutrino as 
additional free parameters. The free transverse momentum 
components $p_{x}^{\nu}$ and $p_{y}^{\nu}$ allow the \pt\ of the 
top-quark pair to float during the fit.
%and the transverse momentum components of the neutrino $p_{x}^{\nu}$ and $p_{y}^{\nu}$, thus allowing the \pt\ of the top-quark pair to float during the fit.

\subsection{The likelihood-based method}
\label{sec:lhmethod}

A likelihood,
\begin{linenomath}
\begin{eqnarray}
\label{eqn:likelihood}
L & =      & B(m_{q_{1}q_{2}q_{3}}|\mtop,\gtop) \cdot  \exp \left( -4 \cdot \ln 2 \cdot \frac{(m_{q_{1}q_{2}} - \mW)^{2}}{\gW^{2}} \right) \nonumber \\ %B(m_{q_{1}q_{2}}|\mW, \gW) \nonumber \\
  & \times & B(m_{q_{4}\ell \nu}  |\mtop,\gtop) \cdot B(m_{\ell \nu}  |\mW, \gW) \nonumber \\
  & \times & \prod_{i=1}^{4} W_{\mathrm{jet}}(E^{\mathrm{meas}}_{\mathrm{jet},i} | E_{\mathrm{jet},i}) \cdot  W_{\mathrm{\ell}}(E^{\mathrm{meas}}_{\ell} | E_{\ell}) \nonumber \\
  & \times & W_{\mathrm{miss}}(\mex | p_{x}^{\nu}) \cdot W_{\mathrm{miss}}(\mey | p_{y}^{\nu}) \, , 
\end{eqnarray}
\end{linenomath}
is maximized for each permutation. The functions $B$ represent
Breit-Wigner functions. The free parameters are the same as for the
$\chi^{2}$-method plus the three momentum components of the
neutrino, corresponding to the second minimization 
option of the $\chi^{2}$-method. The transfer functions are not 
constrained to have a Gaussian shape, see Section~\ref{sec:mc} 
for their parameterization.

% A Gaussian term is used to constrain $m_{q_{1}q_{2}}$ to the mass of 
% the hadronically decaying $W$ boson. Due to the finite energy resolution
% of the jets, the jet energies vary over a broad range of values. 
% As a consequence, a Breit-Wigner function with its dictinct tails accepts large 
% fluctuations of the reconstructed hadronic $W$-boson mass based on 
% these jet energies. Thus, a stronger constraint, as, e.g., provided 
% by a Gaussian term, is chosen.

In a naive leading-order picture, all jets are associated with partons. 
The pole masses of the top quark and the $W$ boson should then be described 
by Breit-Wigner functions. On the stable-particle level, this association between 
jets and partons is not directly possible due to parton showering and hadronization. 
For the studies presented here, we use an approximation for which we replace 
the Breit-Wigner term of the hadronically decaying $W$ boson by a Gaussian 
term as used for the $\chi^{2}$-method in order to constrain $m_{q_{1}q_{2}}$
to the mass of the $W$ boson. 
This approximation is justified by the finite energy resolution of the jets, 
resulting in a variation of the jet energies over a broad range of values. 
As a consequence, a Breit-Wigner function with its distinct tails would accept 
large fluctuations of the reconstructed hadronic $W$-boson mass based on 
these jet energies. Thus, a stronger constraint, as, e.g., provided 
by a Gaussian term, is chosen.
Besides, this procedure leads to the best reconstruction efficiencies found
empirically: a fit using the likelihood-based method with only Breit-Wigner function 
or with only Gaussian terms leads to smaller reconstruction efficiencies than the 
chosen combination of Breit-Wigner and Gaussian terms. Thus other options
are not taken into account for the detailed studies presented in Section~\ref{sec:efficiencies}.

Several extensions to the likelihood defined in
Equation~(\ref{eqn:likelihood}) are available in the current
implementation of this method. These are not used in the direct
comparison of the three reconstruction methods, but the gain in
performance is evaluated separately, see
Section~\ref{sec:performance}. The extensions are:

\paragraph{$b$-tagging} In most cases, $b$-tagged jets -- jets which
are identified to stem from a $b$ quark -- do not originate from the
light quarks of the hadronically decaying $W$ boson. Permutations
which include such an association are vetoed. When considering exactly
four jets in the fit, the number of possible permutations reduces to
six (two) for events with one (two) $b$-tags. Events with more than
two $b$-tagged jets are rejected in the current example because they
are not in agreement with the leading-order interpretation of the
event topology.

\paragraph{Fixed top-quark mass parameter} The mass of the top quark 
is fixed to the value used in the generation of the Monte Carlo
sample. 

\paragraph{Angular information} The decay of the top quark is described
by the $(V-A)$-structure of the electroweak interaction predicting the
angular distribution of the charged lepton. The angle $\theta^{*}$ is
defined as the angle between the charged lepton and the
reverse momentum direction of the top quark, both in the rest frame of
the $W$ boson. The angular distribution is multiplied by the
likelihood,
\begin{linenomath}
\begin{eqnarray}
\frac{1}{\gW}\frac{d\gW}{d\cos(\theta^{*})} & = & 0.687 \cdot \frac{3}{4} \left( 1-\cos^{2}(\theta^{*}) \right)
                   + 0.311 \cdot \frac{3}{8} \left( 1-\cos(\theta^{*}) \right)^{2} \nonumber \\
                   & +&  0.002 \cdot \frac{3}{8} \left( 1+\cos^{2}(\theta^{*}) \right) \,. 
\end{eqnarray}
\end{linenomath}
The numerical coefficients used for the longitudinal, left-handed and
right-handed $W$ boson are those given in
Ref.~\cite{Czarnecki:2010gb}. A corresponding (and symmetrized) term
for the hadronically decaying $W$ boson is also added.

\section{The KLFitter framework}
\label{sec:klfitter}

All three reconstruction methods are implemented in the KLFitter
framework, a C++ program for kinematic fitting based on the Bayesian
Analysis Toolkit ({\it BAT})~\cite{Caldwell:2008fw}. Its modular
structure enables the user to define likelihoods for arbitrary
processes and their corresponding final states -- such as the example
given in Equation~(\ref{eqn:likelihood}) -- and to also specify the
detector model in the form of transfer functions for different
objects. In addition, several minimization and integration algorithms
can be used. The framework provides input and output interfaces to
ASCII and ROOT files, and it can easily be adjusted to any analysis
framework. The code is available on request.

KLFitter has been developed for the case of top-quark reconstruction
and its performance has been studied
extensively~\cite{thesis_fuchs,thesis_ebert,thesis_stolte,thesis_nackenhorst}. It
has been applied in a variety of physics analyses, see
e.g. Refs.~\cite{ATLAS:2012an,ATLAS:2012aj,Aad:2012ky,Aad:2012xc,Aad:2012em,Aad:2012hg,Aad:2013ksa}.
%and it has been extended to final states other than the single-lepton
%decay mode see e.g. Ref.~\cite{ATLAS-CONF-2012-031}.
The application to processes other than top-quark pair production is
straight forward.

\section{Monte Carlo samples and detector modeling}
\label{sec:mc}

The Monte Carlo sample used for the performance studies is generated
using the Sherpa event
generator~\cite{Gleisberg:2008ta,Hoeche:2009rj,Schumann:2007mg} with
the CTEQ6.6~\cite{Nadolsky:2008zw} set of parton distribution
functions and contains 5 million events. The simulated process is top-quark
pair production with up to one additional parton in the final state,
and subsequent decay via the single-electron decay mode.  A
center-of-mass energy of $\sqrt{s}=7$\,TeV is assumed as well as a
top-quark mass of $\mtop=172.5$\,GeV. Particle showering and
hadronization are also performed using Sherpa. Particle jets are built
with FastJet~\cite{Cacciari:2011ma} using the anti-$k_{t}$
algorithm~\cite{Cacciari:2008gp} with a radius parameter of $R=0.4$
and a minimum \pt\ of $15$\,GeV. %It is these particle jets which are
%referred to as final-state particles throughout this paper. 
Those particle jets are referred to as final-state particles 
throughout this paper. 
A matching criterion is added to associate jets at the particle level with 
partons using the distance between these objects defined as
$\Delta R = \sqrt{\Delta \eta^{2} + \Delta \phi^{2}}$,
where $\Delta \eta$ and $\Delta \phi$ are the differences in
pseudo-rapidity and azimuthal angle, respectively.
This distance is required to be smaller than $\Delta R=0.3$. 

For simplicity, detector effects on the measured jets are modeled by
sampling their energies from a Gaussian distribution centered around
the particle-jet energies $E$ [GeV] and with a standard deviation of
$\sigma/\mathrm{GeV}=0.65 \cdot \sqrt{E}/\sqrt{\mathrm{GeV}} + 0.03 \cdot E/\mathrm{GeV}$. For the current example, 
the standard deviation is chosen to be independent of the type of the incident
particle (light or heavy quark, gluon, or hadronically decaying tau
lepton) and its pseudo-rapidity. The numerical values are chosen to
roughly reflect the transfer functions used by the D\O{} 
collaboration~\cite{Abazov:2011ck}. 
The energies of charged leptons are smeared with 
$\sigma/\mathrm{GeV}=0.1 \cdot \sqrt{E}/\sqrt{\mathrm{GeV}}$. 
The missing transverse momentum is defined 
as the negative sum of all matched jet and charged-lepton momenta.

Identification of $b$ jets, \emph{$b$-tagging}, is modeled by choosing
a working point close to what can be found in the literature, see
e.g. Refs.~\cite{Chatrchyan:2012jua,ATLAS:2012aj}. The $b$-tagging
efficiency -- the probability to tag a jet containing a $B$ hadron
-- is assumed to be 70\%, the mis-tag rate -- the probability to
identify a non-$b$ jet as a $b$ jet by mistake -- is assumed to be
0.5\%. Jets are $b$-tagged based on random numbers and on their true
parton flavor. No assumptions are made on the underlying $b$-tag
algorithm or on the output distributions of such an algorithm.

\begin{figure}[ht!]
\begin{center}
\begin{tabular}{cc}
\includegraphics[width=0.48\textwidth]{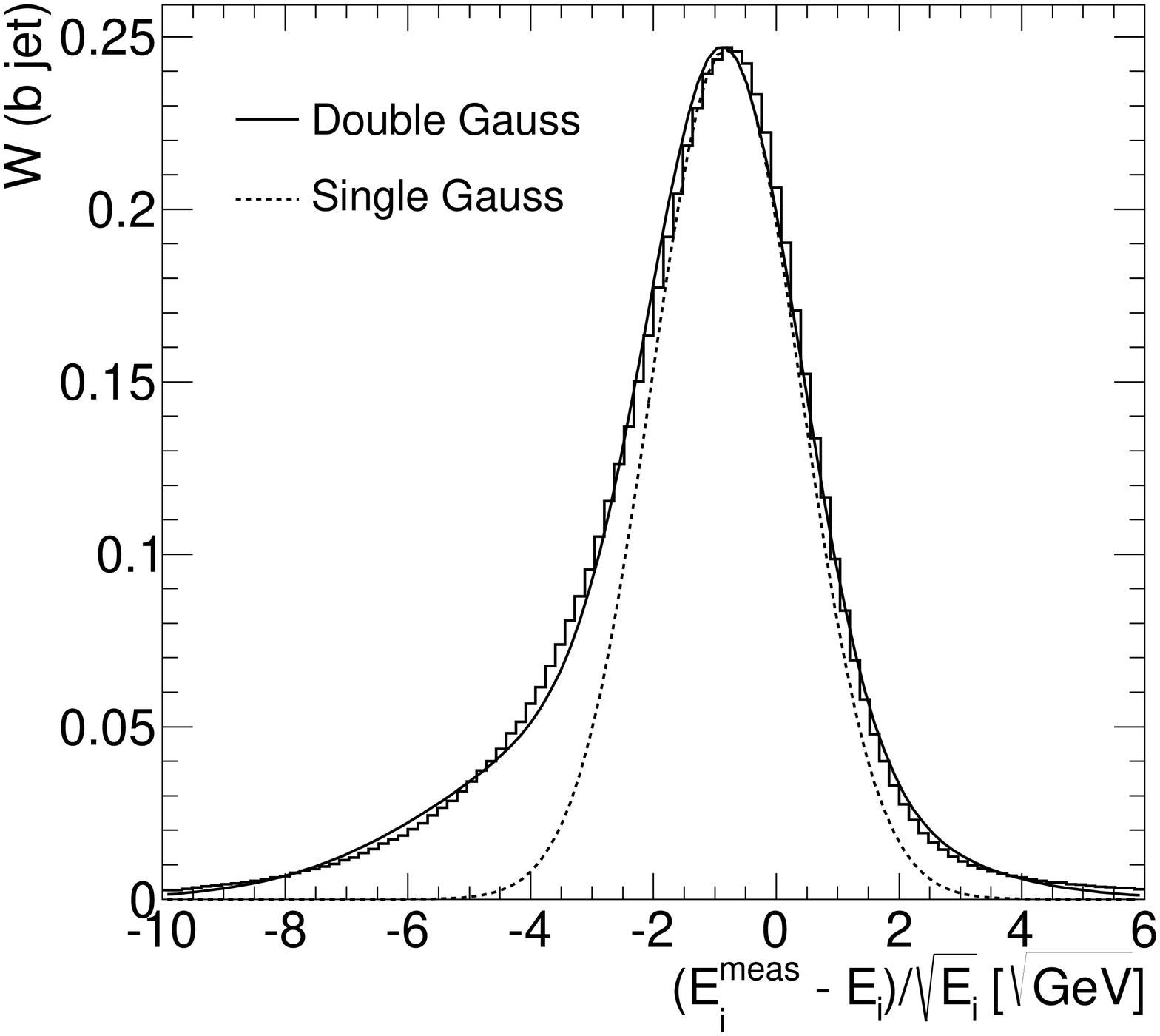} &
\includegraphics[width=0.48\textwidth]{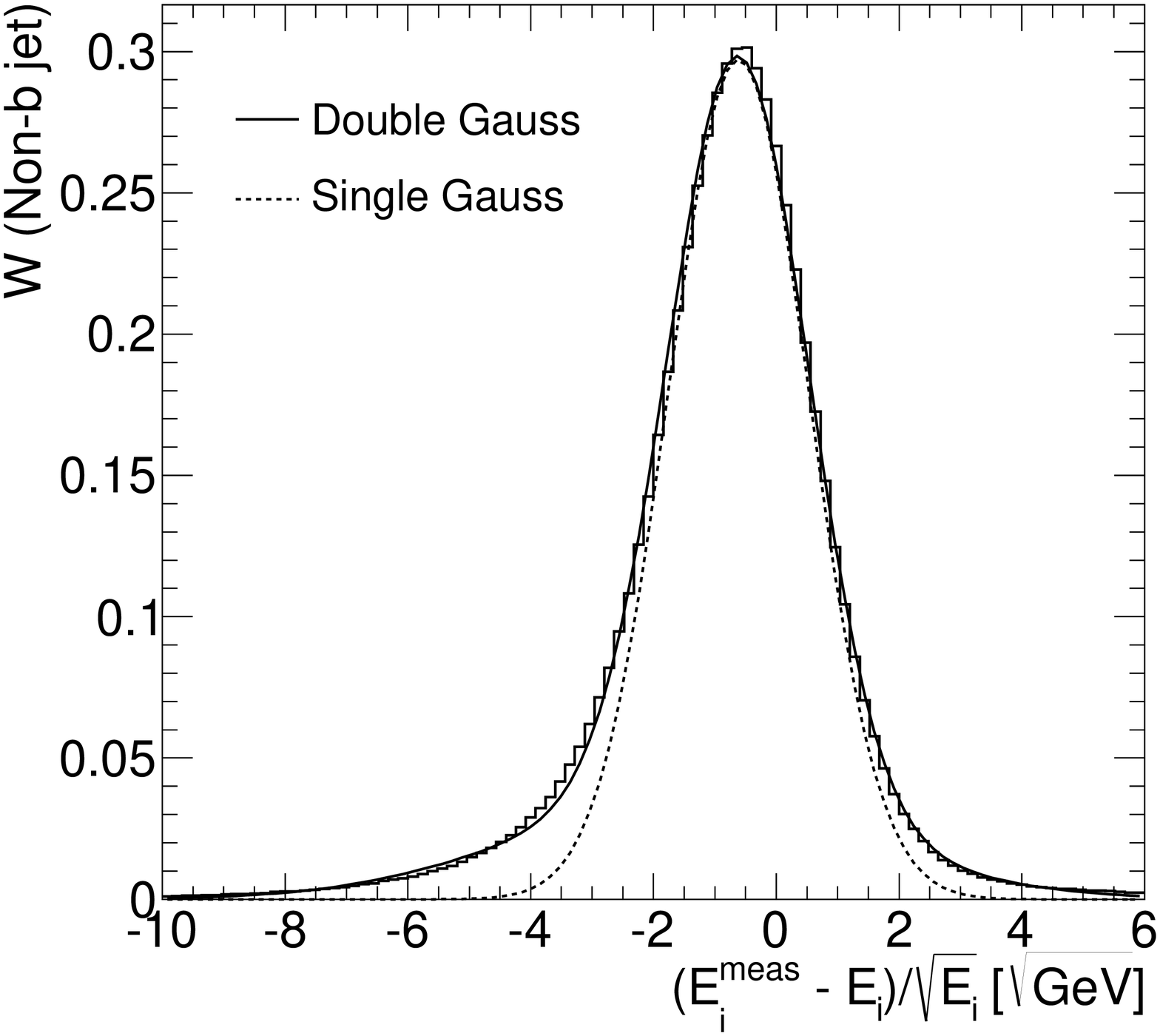} \\
\end{tabular}
\caption{Transfer functions for $b$ jets (left) and non-$b$ jets
  (right) shown as histograms, and the parameterization with a
  double-Gaussian and a single-Gaussian function. Note that, for this
  figure, the normalization of the single-Gaussian function is chosen
  such that it allows for a comparison of shapes.
\label{fig:tf}}
\end{center}
\end{figure}

The transfer functions introduced in Section~\ref{sec:algorithms} are
derived from the differences between the smeared jet energies and the
energy of the particles at the parton level. Figure~\ref{fig:tf} shows two
examples, the distribution of $(E^{\mathrm{meas}}_{i} -
E_{i})/\sqrt{E_{i}}$ for $b$ jets (left) and non-$b$ jets (right). The
transfer functions for the $\chi^{2}$-method and the likelihood-based
method are derived by fitting these distributions with single- and
double-Gaussian functions, respectively. The fit range is constrained
to the peak region for the $\chi^{2}$-method. 
Both parameterizations are indicated in the figure. Section~\ref{sec:discussion} 
includes a discussion on how to implement more realistic transfer functions.
We use the same parameterization for the transfer functions of charged leptons.
On the contrary, transfer functions for the neutrinos are obtained from 
a single-Gaussian fit for both the $\chi^{2}$-method and the 
likelihood-based method.

\section{Event selection}
\label{sec:selection}

The selection of events is motivated by the signature of the
single-lepton decay mode. The requirements for an event to be fitted
are the presence of exactly one electron with $\pt>20$\,GeV and
$|\eta|<2.5$, at least four jets with $\pt>20$\,GeV and $|\eta|<2.5$,
and up to two $b$-tagged jets.

The fraction of events passing the selection, the \emph{selection
efficiency}, is about 19\%. For each event, exactly four jets are
considered in the kinematic fit: first, all $b$ jets are selected in
order of their transverse momentum. Then the remaining
(non-$b$-tagged) jets are added, also sorted by their transverse
momentum.

The set of jets considered in the kinematic fit does not always
correspond to the final-state particles of the hard interaction. For
example, this is due to initial- and final-state radiation, the chosen
jet algorithm and acceptance cuts. 
In a \emph{leading-order picture},
reconstructed jets can be associated with, or \emph{matched} to,
final-state particle jets by requiring their distance in
$(\eta-\phi)$-space to be less than $\Delta R=0.3$. 
%The distance is
%efined as $\Delta R = \sqrt{\Delta \eta^{2} + \Delta \phi^{2}}$,
%where $\Delta \eta$ and $\Delta \phi$ are the differences in
%pseudo-rapidity and azimuthal angle, respectively. 
The \emph{matching efficiency} is defined as the fraction of events 
for which the four jets are unambiguously matched to the final-state particles. 
It is approximately 15\% in the current example and further discussed in
Section~\ref{sec:efficiencies}.

\section{Performance studies}
\label{sec:performance}

The performance of the three reconstruction algorithms is quantified
based on two measures, the reconstruction efficiency and the
reconstruction of top-quark properties. The {\it reconstruction
efficiency} is defined as the fraction of matched events for which the
chosen permutation is indeed the correct one. It is also defined
separately for all objects or a subset of objects, e.g., the
hadronically decaying $W$ boson for which the two corresponding jets
have to be identified correctly. These numbers are compared for the
different reconstruction algorithms in Section~\ref{sec:efficiencies}
as a function of several kinematic quantities and also inclusively. The
object reconstruction is quantified by the distribution of the
difference between the reconstructed objects, e.g., the two top
quarks, and the corresponding \emph{truth} particles (calculated from
the final-state particles). Three classes are compared in
Section~\ref{sec:reconstruction}: matched events for which the chosen
permutation is the correct one, all matched events and all selected
events. The comparison of the first two classes shows the degradation
from the ideal case of a completely efficient reconstruction algorithm
to a less efficient algorithm, both applied to a set of events for
which all selected jets can be associated with the final-state
particles. The comparison of the second and third class shows the
effect of including events for which such an association is not
possible.

\subsection{Reconstruction efficiencies}
\label{sec:efficiencies}

Table~\ref{tab:reco} shows the reconstruction efficiencies for the
three algorithms under study, and those for the modified versions of
the likelihood-based fit. The first three rows give the probability to
identify the correct permutation by chance ignoring any misidentification 
of $b$ jets.

\begin{table}[ht!]
\begin{center}
\begin{footnotesize}
\caption{The reconstruction efficiency for all jets (overall), the
  jets of the hadronically decaying $W$ boson ($W_{\mathrm{had}}$) and
  the two $b$ jets based on events produced at a center-of-mass energy of
  $\sqrt{s} = 7$ TeV. The per-jet probability of correctly identifying a
  $b$-jet and the misidentification rate for light jets are also given.
  The first three rows indicate the reconstruction efficiencies determined 
  from pure combinatorics ignoring misidentification.
  Numbers for the three different reconstruction algorithms without any 
  extensions applied are presented in the next four rows.  
  The further four rows represent these additional options which are 
  subsequently included for the $\chi^{2}$- and the likelihood-based (LH) method. 
  The last three rows show the the reconstruction efficiencies for the 
  likelihood-based method including all extensions split into samples 
  composed of events with zero, one and two $b$-tags. 
  All efficiencies are calculated using matched events.
  The uncertainties are statistical uncertainties. Since the
  different methods are tested on the same data set, these
  uncertainties are correlated.
\label{tab:reco}}
\begin{tabular}{@{}lcccccc@{}}
\toprule
             & \multicolumn{6}{@{}c@{}}{Reconstruction efficiency [\%] }\\  \cline{2-7} 
Method       & Overall & $W_{\mathrm{had}}$ & $b_{\mathrm{had}}$ & $b_{\mathrm{lep}}$ & $p(b\textrm{-id})$ & $p(b\textrm{-mis-id})$ \\
\midrule 
Comb. 0-tag 			& \phantom{1}8.3& \phantom{1}16.7 	& 25.0 & 25.0 & \phantom{1}50.0 & 50.0 \\
Comb. 1-tag 			& 16.7 		& \phantom{1}33.3 	& 33.3 & 33.3 & \phantom{1}66.7 & 33.3 \\
Comb. 2-tag 			& 50.0		& 100.0 		& 50.0 & 50.0 & 100.0 		& \phantom{1}0.0 \\
\midrule 
\ptmax      			& $21.5 \pm 0.1$ & $28.2 \pm 0.2$ & $30.7 \pm 0.2$ & $47.7 \pm 0.2$ & $57.6 \pm 0.2$ & $42.4 \pm 0.2$ \\
$\chi^{2}$ ($m_W$ con.)		& $46.1 \pm 0.2$ & $59.1 \pm 0.3$ & $51.8 \pm 0.3$ & $62.4 \pm 0.3$ & $77.0 \pm 0.3$ & $23.0 \pm 0.2$ \\
$\chi^{2}$  			& $48.4 \pm 0.2$ & $60.8 \pm 0.2$ & $53.8 \pm 0.2$ & $64.8 \pm 0.2$ & $78.0 \pm 0.3$ & $22.0 \pm 0.1$ \\
Likelihood 			& $51.9 \pm 0.2$ & $60.6 \pm 0.2$ & $56.8 \pm 0.2$ & $70.9 \pm 0.2$ & $78.2 \pm 0.3$ & $21.8 \pm 0.1$ \\
\midrule
$\chi^{2}$ +$b$-veto		& $70.7 \pm 0.2$ & $89.1 \pm 0.3$ & $72.7 \pm 0.2$ & $75.5 \pm 0.3$ & $94.6 \pm 0.3$ & $\phantom{1}5.4 \pm 0.1$ \\
LH+$b$-veto  			& $74.3 \pm 0.3$ & $88.8 \pm 0.3$ & $76.4 \pm 0.3$ & $79.5 \pm 0.3$ & $94.4 \pm 0.3$ & $\phantom{1}5.6 \pm 0.1$ \\
~~~~+fix mass			& $83.3 \pm 0.3$ & $91.1 \pm 0.3$ & $84.9 \pm 0.3$ & $88.0 \pm 0.3$ & $95.5 \pm 0.3$ & $\phantom{1}4.5 \pm 0.1$ \\
~~~~+angles    			& $83.8 \pm 0.3$ & $91.2 \pm 0.3$ & $85.3 \pm 0.3$ & $88.4 \pm 0.3$ & $95.6 \pm 0.3$ & $\phantom{1}4.4 \pm 0.1$ \\
\midrule
~~~~+0 $b$-tag			& $\phantom{1}62 \pm 12$ & $\phantom{1}66 \pm 12$ & $\phantom{1}62 \pm 12$ & $\phantom{1}77 \pm 13$ & $\phantom{1}81 \pm 13$ & $19 \pm 7$ \\
~~~~+1 $b$-tag 			& $75.9 \pm 0.4$ & $81.7 \pm 0.4$ & $79.1 \pm 0.4$ & $85.8 \pm 0.4$ & $90.8 \pm 0.4$ & $\phantom{1}9.2 \pm 0.1$ \\
~~~~+2 $b$-tag 			& $90.3 \pm 0.4$ & $99.2 \pm 0.4$ & $90.5 \pm 0.4$ & $90.6 \pm 0.4$ & $99.6 \pm 0.4$ & $\phantom{1}0.4 \pm 0.1$ \\
\bottomrule
\end{tabular}
\end{footnotesize}
\end{center}
\end{table}

All reconstruction efficiencies obtained with the \ptmax-method are
smaller by factors of about $0.25$-$0.5$ compared to those obtained
with the $\chi^{2}$- and likelihood-based methods. This is expected as
the latter two make use of more information of the top-quark decay
topology and of the detector response. In particular, the
\ptmax-method does not use information about the leptonically decaying
top quark resulting in a large misidentification rate for
$b$~jets. The $\chi^{2}$-method with $p_{z}^{\nu}$ as an additional fit 
parameter instead of using a constraint on the $W$-boson mass ($m_W$ con.)   
outperforms the other $\chi^{2}$-method by about 2\% (absolute).
The overall reconstruction efficiency obtained by the
likelihood-based algorithm is about 4\% larger than that 
obtained with the $\chi^{2}$-method. This difference can be
explained by the modeling of the resonant decays (Breit-Wigner
distributions compared to Gaussian distributions) and the more
accurate parameterization of the transfer functions. %, see Figure~\ref{fig:tf}.

Using a veto for $b$-tagged jets which are interpreted as originating
from light quarks results in an increase in the reconstruction
efficiencies. In particular, the $b$-misidentification rate decreases
significantly and, as a consequence, the probability to correctly
identify the hadronically decaying $W$~boson increases. 
This applies to both the $\chi^{2}$- and the likelihood-based method for 
which this $b$-veto is tested. The $\chi^{2}$-method in combination with
a veto for $b$-tagged jets is similar to the reconstruction method 
used in Ref.~\cite{CMS:2012tm}, where only events with two $b$-tags are used. 
As expected, the overall reconstruction efficiency
for the likelihood-based method is better by 4\% and thus the remaining
extensions are studied for this method only. Additionally fixing the
top-quark mass to the value used in the simulated samples increases
the reconstruction efficiencies even further due to the tighter
constraints on the jet energies in comparison to the Breit-Wigner functions. 
The addition of angular information increases the efficiencies only
slightly.

The presented numbers in the last three rows of Table~\ref{tab:reco} are based 
on a likelihood fit using all possible extensions and show that the 
reconstruction efficiency depends strongly on the number
of $b$-tagged jets in an event. The reconstruction efficiencies rise
from 62\% for events without any $b$-tag (corresponding to a fraction of
0.03\% of all matched events) to 75.9\% (90.3\%) for events with exactly one
(two) $b$-tagged jets (corresponding to fractions of 46\% and 54\% of all
matched events, respectively).

\begin{figure}[ht!]
\begin{center}
\begin{tabular}{cc}
\includegraphics[width=0.48\textwidth]{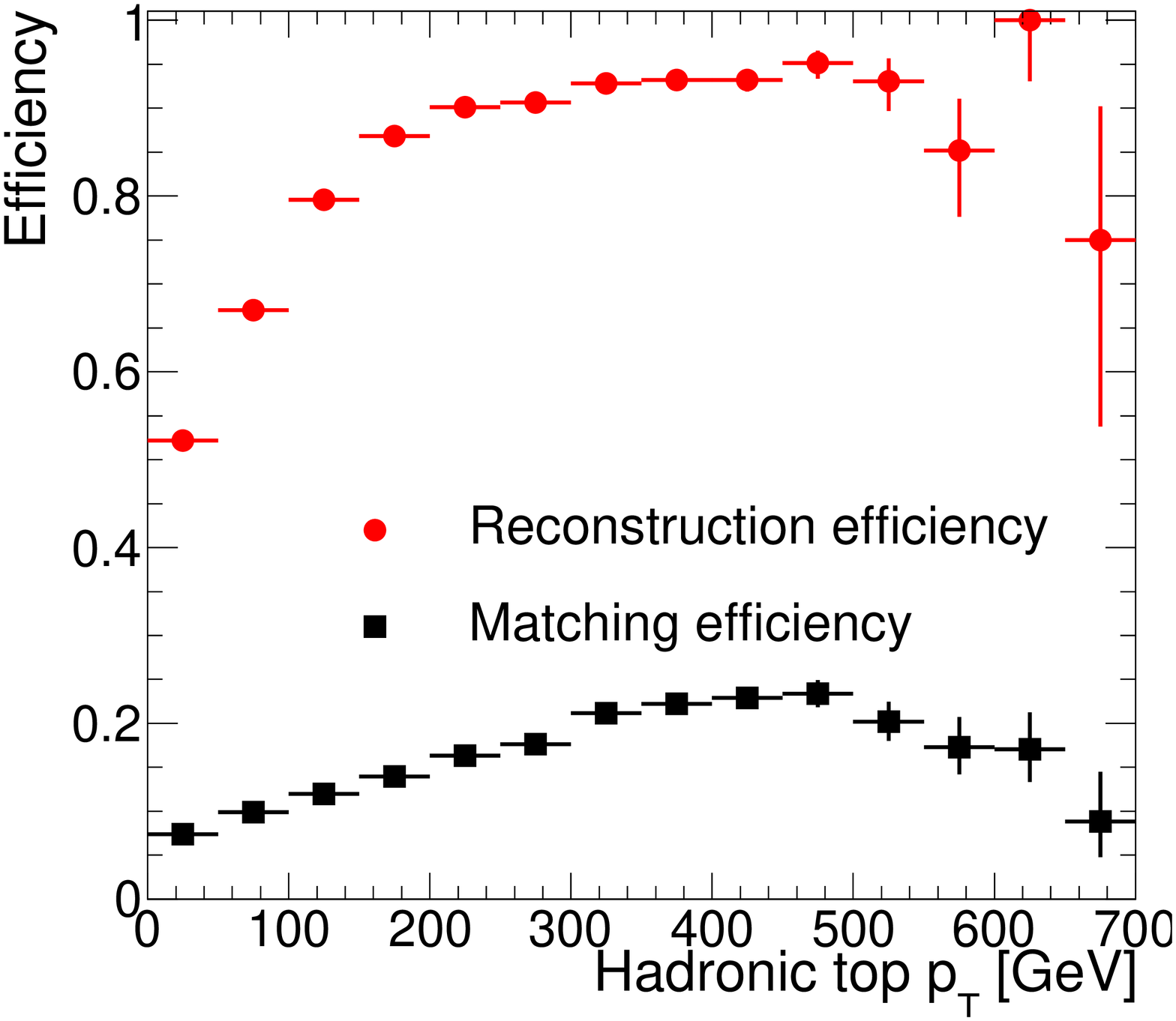} &
\includegraphics[width=0.48\textwidth]{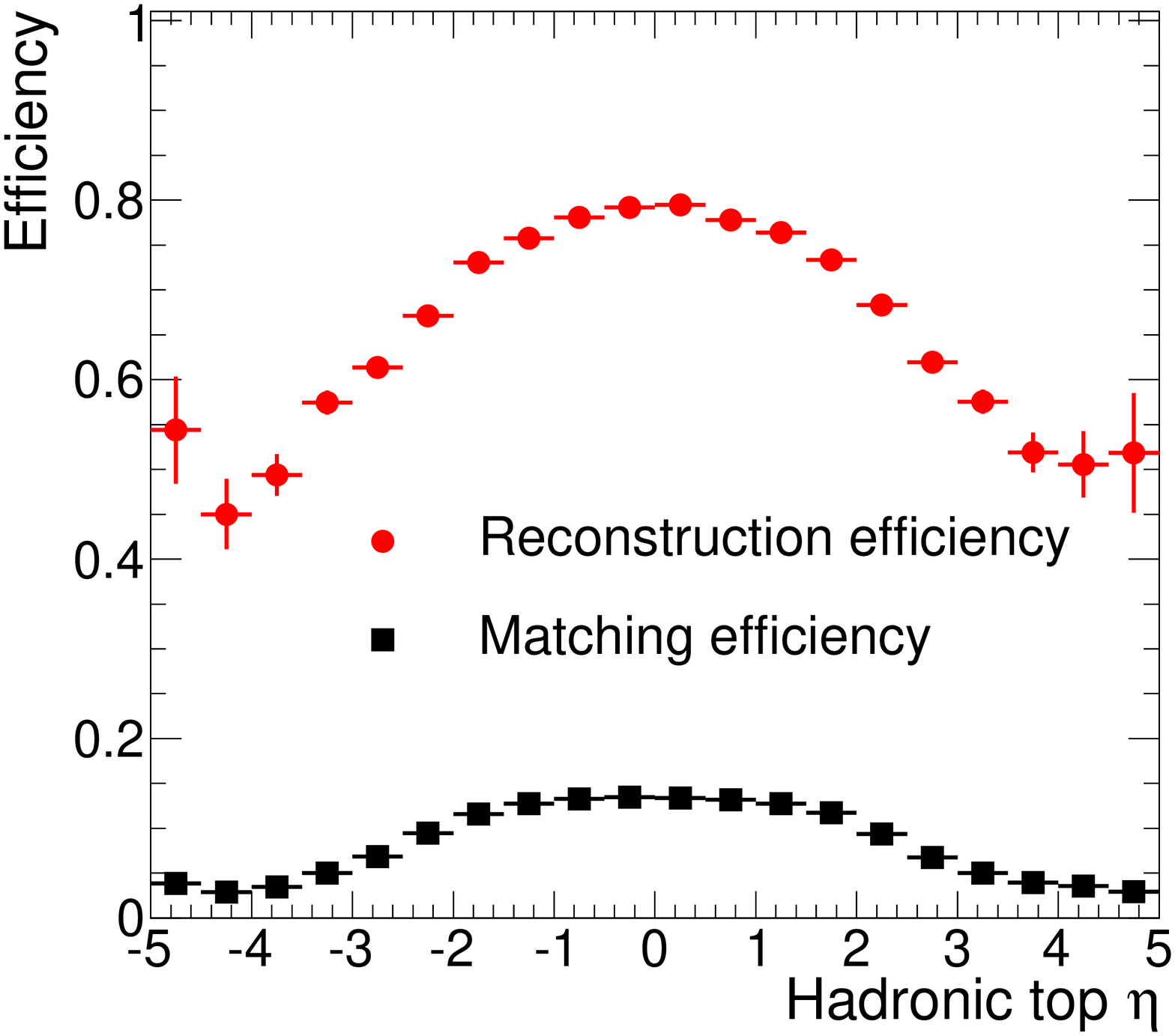} \\
\end{tabular}
\caption{Matching and (overall) reconstruction efficiency for the
  likelihood-based reconstruction algorithm with the additional $b$-tag veto
  as a function of the
  transverse momentum (left) and the pseudo-rapidity (right) of the
  hadronically decaying top quark.
\label{fig:eff}}
\end{center}
\end{figure}

Figure~\ref{fig:eff} (left) shows the matching efficiencies, as
defined as the fraction of events for which the four jets are 
unambiguously matched to the final-state particles, and (overall)
reconstruction efficiencies for the likelihood-based algorithm
(including the $b$-tag veto) as a function of the transverse momentum
of the hadronically decaying top quark. The matching efficiency
increases from about 10\% at low \pt\ to roughly 25\% at $\pt \approx
500$\,GeV, and it drops again for larger values. 
This drop can be explained by the fact that the decay
products of the top quark are highly boosted and, in many cases, 
cannot be resolved by
the jet algorithm. The reconstruction efficiency increases from about
50\% at low \pt\ to about 90\% at $\pt \approx 250$\,GeV and then
remains constant for higher \pt\ values. For large transverse momenta,
it is less likely that a wrong combination of jets results in an
invariant mass close to that of the top quark because the jets are
strongly collimated compared to the case of low \pt\ values. Although
high-\pt\ top quarks can be reconstructed rather well, the fraction of
events in which four resolved jets on the reconstruction level correspond
to the particle-level jets is small.

Figure~\ref{fig:eff} (right) shows the two efficiencies as a function
of the pseudo-rapidity of the hadronically decaying top quark. Both
efficiencies are approximately constant in the central region,
i.e. $|\eta|<1.5$, and drop for larger values of $|\eta|$. This drop
can be explained by the fact that the decay products of the top quark
are more likely to be found outside the acceptance region of
$|\eta|<2.5$.

\subsection{Object reconstruction}
\label{sec:reconstruction}

\begin{figure}[ht!]
\begin{center}
\begin{tabular}{cc}
\includegraphics[width=0.48\textwidth]{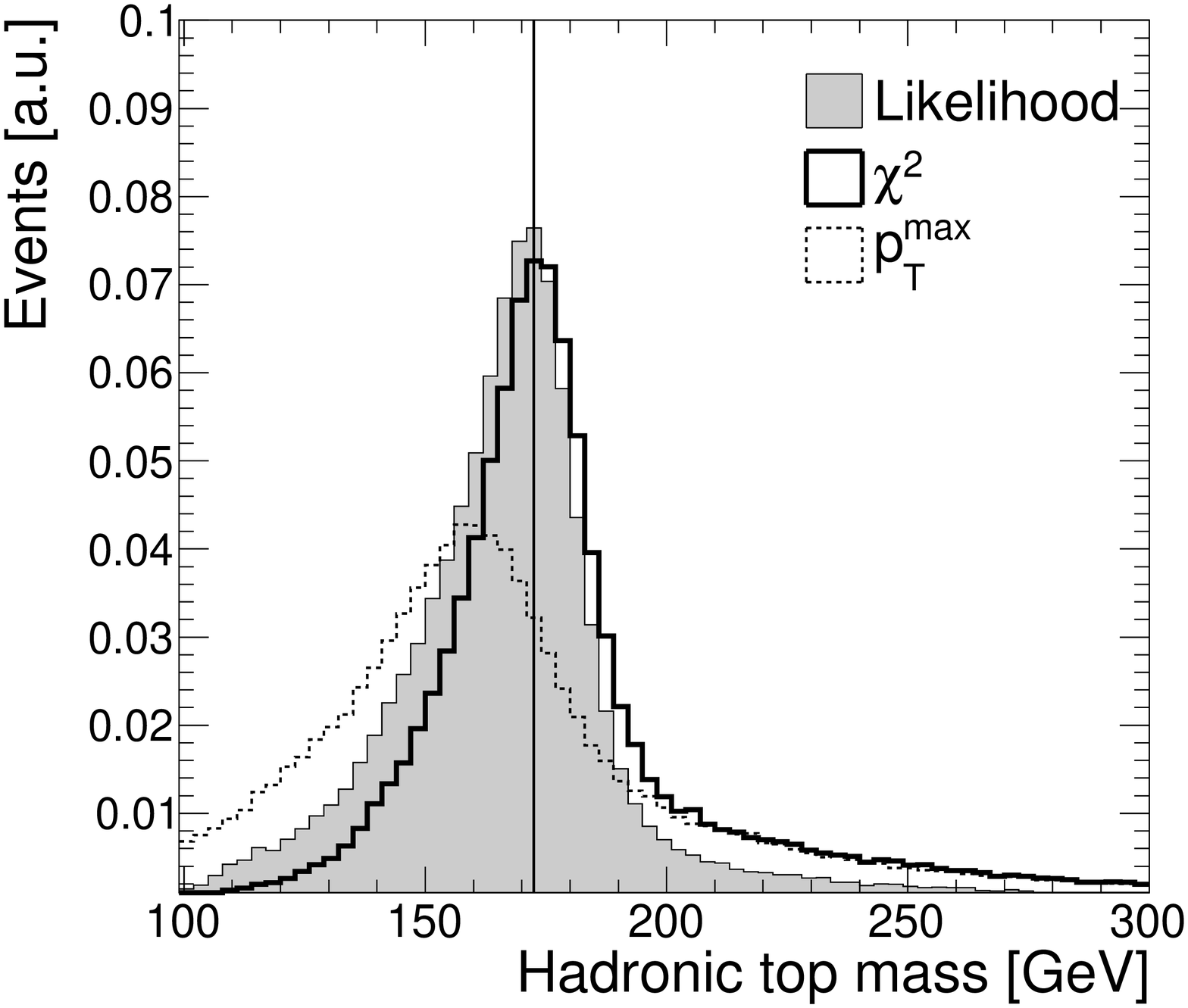} &
\includegraphics[width=0.48\textwidth]{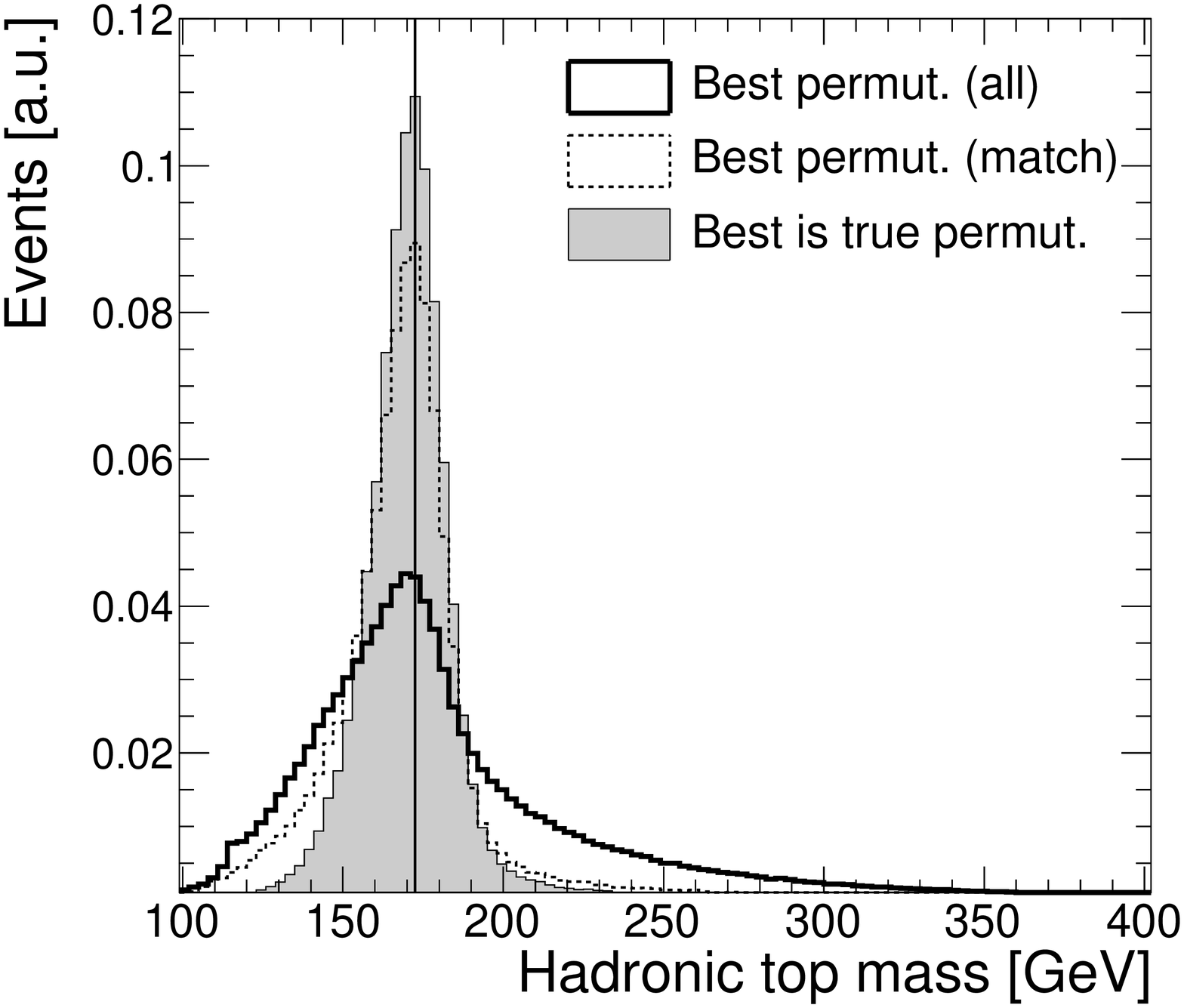} \\
\end{tabular}
\caption{Mass of the hadronically decaying top quark reconstructed
  with the three different reconstruction algorithms using matched 
  events (left) and with the likelihood-based algorithm including the
  $b$-tag veto (right) for three different subsets of the Monte Carlo
  data sample. The black line indicates the top-quark mass (physical parameter)
  assumed for the generation of Monte Carlo events.
\label{fig:obj1}}
\end{center}
\end{figure}

Figure~\ref{fig:obj1} (left) shows the mass of the hadronically
decaying top quark reconstructed with the three different algorithms
using the full selected Monte Carlo data set. This mass is calculated using
the invariant masses obtained from the jet energies. The distribution
obtained by the \ptmax-method is the broadest and peaks at around
$160$\,GeV. The distributions obtained from the $\chi^{2}$- and the
likelihood-based methods are significantly narrower and peak closer to
the top-quark mass assumed in the simulation. The tail to large masses
is more pronounced for the $\chi^{2}$-reconstruction. For this comparison,
the $\chi^{2}$-method treating the momentum components of the neutrino 
as free parameters is used.

The right-hand side of Figure~\ref{fig:obj1} shows the same mass, but
reconstructed with the likelihood-based algorithm (including the
$b$-tag veto). Three different subsets of the Monte Carlo data set are
considered: all events in the sample, matched events and matched
events for which the correct permutation is chosen by the
algorithm. The three distributions are normalized to unity. 
The latter distribution is
approximately symmetric around the input mass while the distribution
of all matched events has slightly larger tails due to the
combinatorial background. The distribution for all events has a
significant tail to larger mass values due to the combinatorial
background and events in which not all final-state particles can be
matched to the selected jets.

Thus, two effects illustrate the impact of a more sophisticated 
reconstruction algorithm characterized by increased reconstruction 
efficiencies:
the amount of combinatorial background is reduced since fewer top quarks 
are wrongly reconstructed and detector effects are reduced. Both effects 
lead to a smaller and more distinct peak in the distribution of the
reconstructed top-quark mass as shown in Figure~\ref{fig:obj1}.

\begin{figure}[ht!]
\begin{center}
\begin{tabular}{cc}
\includegraphics[width=0.48\textwidth]{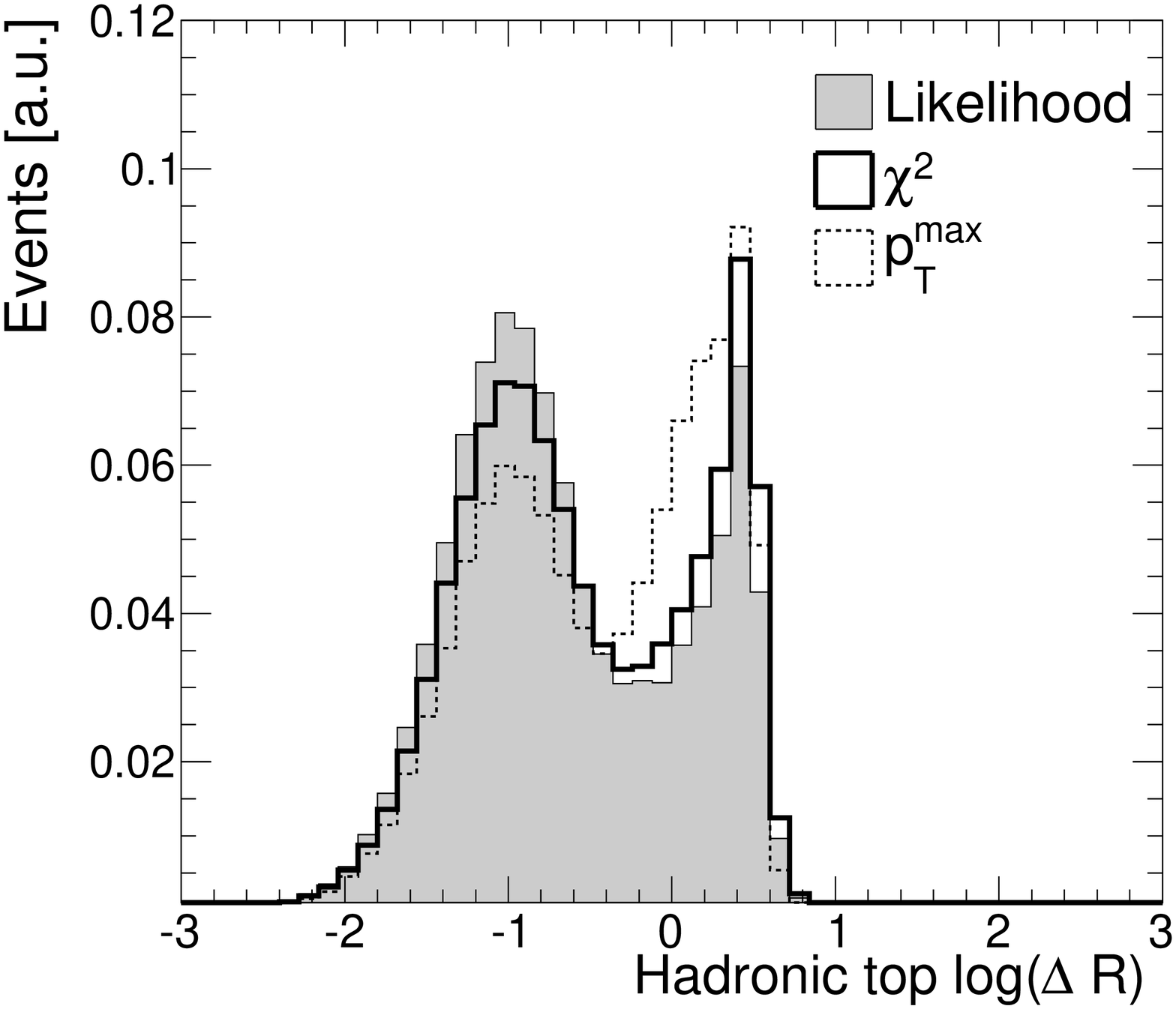} &
\includegraphics[width=0.48\textwidth]{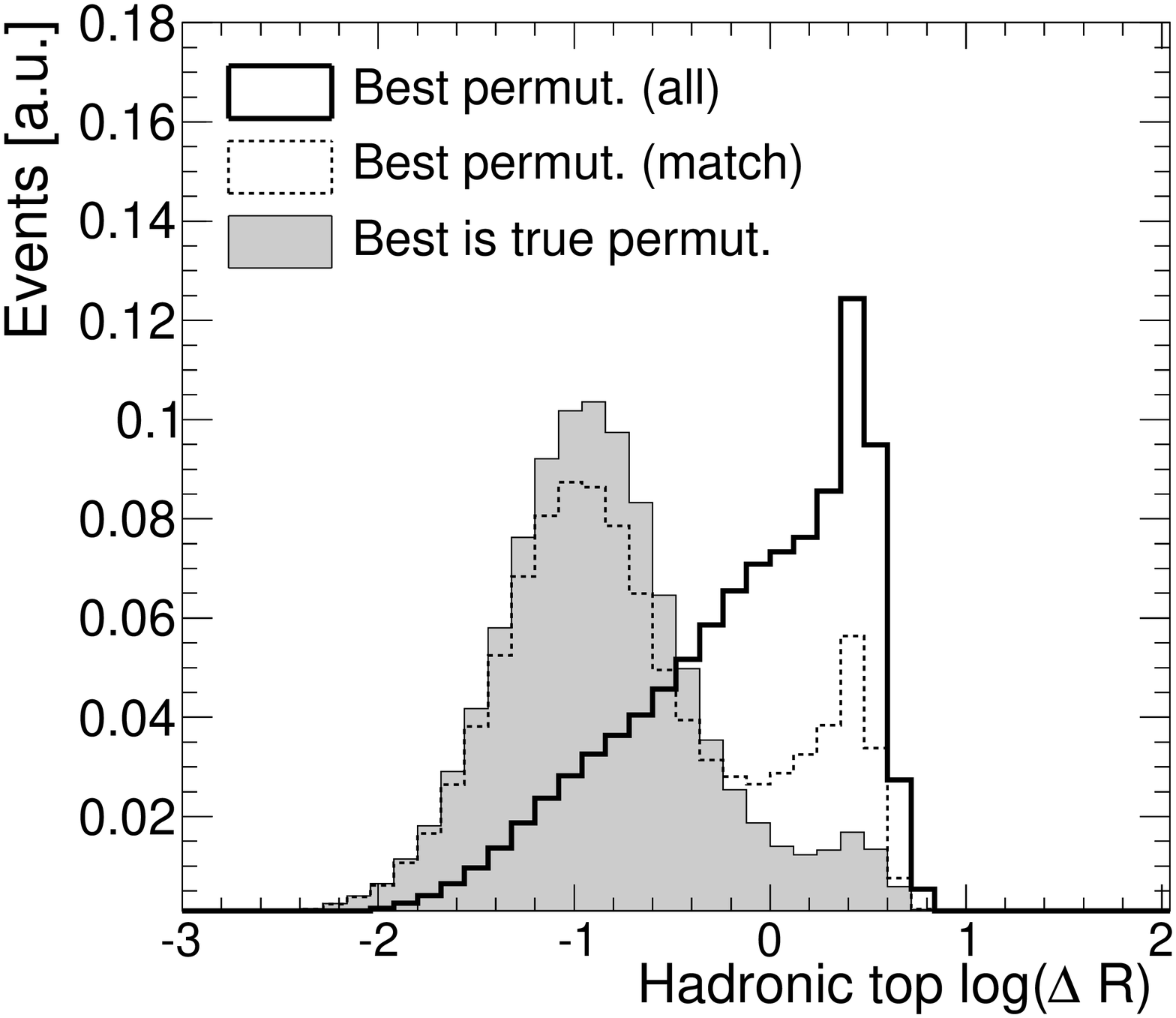} \\
\end{tabular}
\caption{The distance $\log \Delta R$ between the (true) hadronically decaying top quark
  and the one reconstructed with the three different reconstruction
  algorithms using matched events (left) and with the likelihood-based 
  algorithm including the $b$-tag veto (right) for three different subsets
  of the Monte Carlo data sample.
\label{fig:obj2}}
\end{center}
\end{figure}

Figure~\ref{fig:obj2} (left) shows the $\log \Delta R$ between the
(true) hadronically decaying top quark and the reconstructed top quark
for the three different reconstruction algorithms using only matched events. 
The distributions
for all three algorithms show two peaks, one around $\log \Delta R
\approx -1$ and one around $\log \Delta R \approx 0.5$. The latter is
due to combinatorial background while the former reflects the
resolution with which particle-level top quarks can be matched to
reconstructed top quarks. As expected, the peak at smaller values is
more pronounced for the likelihood-based
approach. Figure~\ref{fig:obj2} (right) shows $\log \Delta R$ for
events reconstructed with the likelihood-based algorithm including the
$b$-tag veto for three different subsets of the Monte Carlo data
sample. Matched events for which the chosen permutation corresponds to
the true one have a peak around $\log \Delta R \approx -1$. The
distribution of all events raises continuously and peaks at $\log
\Delta R \approx 0.5$ due to the quasi-random assignment of jets to
the hadronically decaying top quark.

\section{Discussion}
\label{sec:discussion}

\subsection{Summary}
\label{sec:summary}

We presented a flexible framework for kinematic fitting which can be
applied to arbitrary event topologies and models. For the performed studies, 
a likelihood-based reconstruction algorithm was introduced and its
performance was compared to that of two commonly used algorithms for
the case of top-quark pair production. It was found that the
likelihood-based algorithm results in the largest reconstruction
efficiency because of a more accurate description of the underlying
physics process and detector effects.

Although this algorithm performs best at moderate to large transverse
top-quark momenta, the matching efficiency drops in that regime
indicating that a different class of algorithms has to be used to
reconstruct boosted top-quark pairs, see
e.g. Refs.~\cite{Kaplan:2008ie,Thaler:2008ju,Plehn:2010st,Aad:2012raa,Chatrchyan:2012ku}.

If more than four jets are considered in the fit, the number of ways
to associate four jets out of the selected ones with the final-state
particles increases. This increases the matching efficiency because
the probability to select those jets corresponding to the final-state
particles increases. On the other hand, the reconstruction efficiency
decreases because it becomes more likely to fulfill the constraints in
Equation~\ref{eqn:likelihood} by chance for permutations with a wrong
assignment of jets, see Ref.~\cite{thesis_nackenhorst}.

Different types of likelihood-based reconstruction algorithms in KLFitter
have been successfully applied in a variety of physics analyses.

\subsection{Future developments}
\label{sec:developments}

The transfer functions used by the $\chi^{2}$ and likelihood-based
algorithms are assumed to not depend on the particle's properties, and
only a rough distinction is made between light and $b$ jets. 
In a real experiment, however, the energy
measured in the calorimeter indeed depends on those
specifications. For example, jets originating from light quarks or
gluons will differ in their shape and visible energy compared to jets
containing $B$ hadrons. The latter also might need to be categorized
according to the decay of the $B$ hadron because leptonic decays of
$B$ mesons result in neutrinos which carry away some of the energy of
the jet. Furthermore, the quality of the jet reconstruction may depend
on the pseudo-rapidity as different detector components may be used
for the reconstruction. More realistic transfer functions can also be
obtained by considering the energy dependence of the parameterization of 
these functions. Apart from that, further studies may investigate the 
impact of pile-up effects on the matching and the reconstruction 
efficiencies.

Future developments also include a refinement of the likelihood using,
e.g., the leading-order matrix element for top-quark pair production and
decay or the pole mass distributions from the $W$ boson and the top quark 
as obtained from truth Monte Carlo.

%Despite these possible developments, the KLFitter has already been proven
%to be a very performant tool and a valuable contribution to physics analyses
%at the LHC.  

\subsection{Outlook}
\label{sec:outlook}

As an outlook to the running of the LHC with an increased
center-of-mass energy and in view of the increased pile-up, 
the studies are repeated with Monte Carlo simulations assuming 
a center-of-mass energy of $\sqrt{s}=14$\,TeV. 
The same transfer functions are used as for the 7 TeV study.

\begin{table}[ht!]
\begin{center}
\begin{footnotesize}
\caption{The reconstruction efficiency for all jets (overall), the
  jets of the hadronically decaying $W$ boson ($W_{\mathrm{had}}$) and
  the two $b$ jets based on events produced at a center-of-mass energy of
  $\sqrt{s} = 14$ TeV. The per-jet probability of correctly identifying a
  $b$-jet and the misidentification rate for light jets are also given.
  The first three rows indicate the reconstruction efficiencies determined 
  from pure combinatorics ignoring misidentification.
  Numbers for the three different reconstruction algorithms without any 
  extensions applied are presented in the next four rows.  
  The further four rows represent these additional options which are 
  subsequently included for the $\chi^{2}$- and the likelihood-based (LH) method. 
  The last three rows show the the reconstruction efficiencies for the 
  likelihood-based method including all extensions split into samples 
  composed of events with zero, one and two $b$-tags. 
  All efficiencies are calculated using matched events.
  The uncertainties are statistical uncertainties. Since the
  different methods are tested on the same data set, these
  uncertainties are correlated.
\label{tab:reco14TeV}}
\begin{tabular}{@{}lcccccc@{}}
\toprule
             & \multicolumn{6}{@{}c@{}}{Reconstruction efficiency [\%] }\\  \cline{2-7} 
Method       & Overall & $W_{\mathrm{had}}$ & $b_{\mathrm{had}}$ & $b_{\mathrm{lep}}$ & $p(b\textrm{-id})$ & $p(b\textrm{-mis-id})$ \\
\midrule 
Comb. 0-tag 			& \phantom{1}8.3& \phantom{1}16.7 	& 25.0 & 25.0 & \phantom{1}50.0 & 50.0 \\
Comb. 1-tag 			& 16.7 		& \phantom{1}33.3 	& 33.3 & 33.3 & \phantom{1}66.7 & 33.3 \\
Comb. 2-tag 			& 50.0		& 100.0 		& 50.0 & 50.0 & 100.0 		& \phantom{1}0.0 \\
\midrule 
\ptmax      			& $23.6 \pm 0.2$ & $29.8 \pm 0.2$ & $31.9 \pm 0.2$ & $52.4 \pm 0.2$ & $59.2 \pm 0.2$ & $40.8 \pm 0.2$ \\
$\chi^{2}$ ($m_W$ con.)		& $48.1 \pm 0.3$ & $61.5 \pm 0.3$ & $53.2 \pm 0.3$ & $64.3 \pm 0.3$ & $78.5 \pm 0.3$ & $21.5 \pm 0.2$ \\
$\chi^{2}$  			& $50.0 \pm 0.3$ & $62.7 \pm 0.3$ & $54.8 \pm 0.3$ & $66.4 \pm 0.3$ & $79.2 \pm 0.3$ & $20.8 \pm 0.2$ \\
Likelihood  			& $55.0 \pm 0.3$ & $62.7 \pm 0.3$ & $59.3 \pm 0.3$ & $74.5 \pm 0.3$ & $79.5 \pm 0.3$ & $20.5 \pm 0.2$ \\
\midrule
$\chi^{2}$ +$b$-veto		& $71.6 \pm 0.3$ & $89.9 \pm 0.3$ & $73.4 \pm 0.3$ & $76.1 \pm 0.3$ & $94.9 \pm 0.3$ & $\phantom{1}5.1 \pm 0.1$ \\
LH+$b$-veto  			& $77.1 \pm 0.3$ & $89.7 \pm 0.3$ & $78.9 \pm 0.3$ & $82.2 \pm 0.3$ & $94.8 \pm 0.3$ & $\phantom{1}5.2 \pm 0.1$ \\
~~~~+fix mass  			& $85.1 \pm 0.3$ & $92.0 \pm 0.3$ & $86.5 \pm 0.3$ & $89.6 \pm 0.3$ & $96.0 \pm 0.3$ & $\phantom{1}4.0 \pm 0.1$ \\
~~~~+angles    			& $85.6 \pm 0.3$ & $92.1 \pm 0.3$ & $86.9 \pm 0.3$ & $90.0 \pm 0.3$ & $96.1 \pm 0.3$ & $\phantom{1}3.9 \pm 0.1$ \\
\midrule
~~~~+0 $b$-tag 			& $\phantom{1}72 \pm 15$ & $\phantom{1}75 \pm 16$ & $\phantom{1}75 \pm 16$ & $\phantom{1}88 \pm 17$ & $\phantom{1}84 \pm 16$ & $16 \pm 7$ \\
~~~~+1 $b$-tag 			& $78.6 \pm 0.4$ & $83.7 \pm 0.4$ & $81.2 \pm 0.4$ & $87.9 \pm 0.5$ & $91.8 \pm 0.4$ & $\phantom{1}8.2 \pm 0.1$ \\
~~~~+2 $b$-tag 			& $91.4 \pm 0.4$ & $99.2 \pm 0.4$ & $91.6 \pm 0.4$ & $91.8 \pm 0.4$ & $99.6 \pm 0.4$ & $\phantom{1}0.4 \pm 0.1$ \\
\bottomrule
\end{tabular}
\end{footnotesize}
\end{center}
\end{table}

The selection efficiency of this sample amounts to about 16\%. As
the average transverse momentum of the top quarks increases, the
matching efficiency decreases to roughly 13\%. On the other hand, and
consistent with Figure~\ref{fig:eff} (left), the different
reconstruction efficiencies for matched events increase by about
$2-3$\%. The corresponding numbers for the tested reconstruction algorithms
are listed in Table~\ref{tab:reco14TeV}. Likelihood-based fitting will hence 
also be valuable for future measurements at the LHC.

%This behaviour indicates that the likelihood-based fitting will be
%valuable for future measurements at the LHC.

\section*{Acknowledgements}

We would like to thank all users of KLFitter who helped to improve the
package.
%, in particular Stefanie Adomeit, Thomas Loddenk{\"o}tter,
%Leonid Serkin and Tamara Vazquez-Schr{\"o}der for the implementation
%of likelihoods for decay modes other than the single-lepton decay
%mode.
We are also very thankful to Steffen Schumann for his support in the
event generation with Sherpa.

\newpage



%% References


\bibliographystyle{model1a-num-names}
\bibliography{bibliography}






% \end{thebibliography}

\end{document}